\documentclass{aa}  

\usepackage{longtable}
\usepackage{graphicx}
\usepackage{txfonts}
\usepackage{longtable,lscape}
\usepackage{booktabs} 

\def\sun{\hbox{$\odot$}}
\def\R23{\mbox{$\rm R_{23}$}}

\def\kmsmpc{km s$^{-1}$ Mpc$^{-1}$}

\def\Hb{\mbox{${\rm H}{\beta}$}}
\def\Ha{\mbox{${\rm H}{\alpha}$}}
\begin{document}

\title{Star-formation quenching  of cluster galaxies as traced by metallicity, presence of active galactic nuclei, and galactic conformity}


\author{C.~Maier\inst{1}
\and C.\,P.\,~Haines\inst{2}
\and B.\,L.~Ziegler\inst{1}
}

\institute{University of Vienna, Department of Astrophysics, Tuerkenschanzstrasse 17, 1180 Vienna, Austria\\
\email{christian.maier@univie.ac.at}
\and Instituto de Astronomía y Ciencias Planetarias de Atacama, Universidad de Atacama, Copayapu 485, Copiapo, Chile
}

\titlerunning{Galactic conformity in clusters}
\authorrunning{C. Maier et al.}

\date{Received ; accepted}

\abstract 
{}
{
  We strive to explore the differences in the properties and quenching processes of satellite galaxies in a sample of massive clusters with passive and star-forming (SF) brightest cluster galaxies (BCGs). One aim is to investigate galactic conformity effects, manifested in a correlation between the fraction of satellite galaxies that halted star formation  and the state of star formation in the central galaxy.
}
{
  We explored 18
  clusters from the Local Cluster Substructure Survey (LoCuSS) at $0.15<z<0.26,$ using spectra from the ACReS (Arizona Cluster Redshift Survey) Hectospec survey of about 1800
  cluster members at $R<R_{200}$ in a mass-complete sample.
  Nine clusters have a SF BCG and nine have a passive BCG, which enable the exploration of galactic conformity effects.
We measured the fluxes of
emission lines  of cluster members, allowing us to
derive O/H gas metallicities and to identify active galactic nuclei (AGN).
We compared our cluster galaxy sample with a control field sample of about 1300
galaxies with similar masses and at similar redshifts observed with Hectospec as part of the same survey.
We used the location of
SF galaxies, recently quenched galaxies (RQGs) and AGN in the projected velocity versus the position phase-space (phase-space diagram) to identify objects in the inner regions of the clusters and to compare their fractions
in clusters with SF and passive BCGs.
}
{
The metallicities of  log(M/M$_{\sun}) \geq 10$ SF cluster galaxies with $R<R_{200}$ were found to be enhanced with respect to the mass-metallicity
relation obtained for our sample of coeval field SF galaxies.
This metallicity enhancement among SF cluster galaxies is limited to lower-mass satellites ($10 < \rm{log(M/M}_{\sun}) < 10.7$) of the nine clusters with a passive BCG, with no metallicity enhancement seen for SF galaxies in clusters with
active BCGs. Many of the SF galaxies with enhanced metallicities are found in the
core regions of the phase-space diagram expected for virialized populations.
We find a higher fraction of log(M/M$_{\sun}) \geq 10.7$ SF galaxies at $R<R_{500}$ in clusters with active BCGs as compared to clusters with passive BCGs, which stands as a signal of galactic conformity.
In contrast, much higher fractions of AGN, and especially RQGs at $R<R_{500}$, are found in clusters with passive BCGs in comparison to clusters with active BCGs.
}
{
We deduce that strangulation is initiated in clusters with passive BCGs when SF satellite galaxies pass $R_{200}$, by stopping the pristine gas inflow that would otherwise dilute the interstellar medium and would keep their metallicities at the level of values similar to those of field galaxies at similar redshifts.  
These satellite galaxies continue to form stars by consuming the available gas in the disk.
For galaxies with massses above log(M/M$_{\sun}) \sim 10.7$ that manage to survive and remain SF when traveling to $R<R_{500}$ of clusters with passive BCGs, we assume that they suffer a rapid quenching of star formation, likely due to AGN triggered by the increasing ram pressure stripping toward the cluster center, which can compress the gas and fuel AGN.
  These AGN can rapidly quench and maintain quenched satellite galaxies.  On the other hand,  we found that surviving SF massive satellite galaxies around active BCGs are less affected by environment when they enter $R<R_{500}$, since we observe  $R<R_{500}$ SF galaxies with masses up to $M  \sim 10^{11}M_{\odot} $ and with  metallicities typical of coeval field galaxies. This observed galactic conformity implies that active BCGs must maintain their activity over timescales of at least $\sim 1$\,Gyr.
}



\maketitle



\setcounter{section}{0}
\section{Introduction}
\label{sec:intro}

~~~One of the key questions in galaxy evolution that has arisen in recent years considers whether galaxies "conform" to their environment, namely: whether the star formation rates (SFR) of satellite galaxies are coupled to the SFR of the brightest galaxy in their group or cluster; whether the satellite galaxies that reside only in the inner regions of groups or clusters are affected or also in regions further outside; and  whether there are hints about the physical origin of such a "galactic conformity"\ that can be identified. A general question in this context is whether the suppression of star formation (quenching) in satellite and central galaxies is connected.

~~~The specific star formation rate (sSFR = SFR$/M_{*}$) has been found to be a tight but weak function of stellar mass at all epochs up to redshift $z \sim 2$ for most star-forming (SF) galaxies (on the main sequence), and to decrease with cosmic time being substantially elevated,
by a factor of about 20 at $z=2$ and by about 7 at $z=1$ compared to $z \sim 0$ \citep[e.g.,][]{elbaz07,peng10}.
Some processes that suppresses star-formation in galaxies to a level $1-2$ dex lower (or more), relative to the level in main sequence galaxies of the same mass, is described as ``quenching.'' It transforms ``active'' SF galaxies that are forming stars at a cosmologically significant rate, $sSFR\ge \tau_{H}^{-1}$ (with  $\tau_{H}$ the Hubble time) into ``passive'' galaxies that are not forming stars at a cosmologically significant rate, specifically, with $sSFR\ll \tau_{H}^{-1}$ \citep[see, e.g., quenching paths for galaxies in Fig.\,4 in][]{maier09}.


~~~ The fraction of passive galaxies has been measured as a function of stellar mass and environment in the local Universe \citep[e.g.,][]{kauf04,bald06,haines07}  and at high redshift \citep[e.g.,][]{patel09,muzzin12}. These studies revealed that quenching
can be divided into two flavors: a "mass-dependent" and an "environment-dependent" mode \citep[e.g.,][]{peng10}.
Numerous physical processes have been proposed in the literature to explain the cessation of star formation, such as halo quenching \citep[e.g.,][]{zumandel16} and AGN quenching \citep[e.g.,][]{schaye15} for ``mass-quenching,'' as well as strangulation \citep[e.g.,][]{peng15,maier16} and ram pressure stripping \citep[RPS, e.g.,][]{bahe13,poggianti17,maier19a,maier19b} for ``environment-quenching.''

~~~ Mass-quenching is the dominant quenching mechanism affecting central galaxies, the most massive galaxies in their halo residing at the potential minimum. Environment-quenching should be more pronounced on satellite galaxies, which are moving relative to the potential minimum having fallen into the larger halo. However, there is some evidence of a connection between the halting of star formation in central and satellite galaxies.
This particular phenomenon, which has gained attention in recent years is so-called ``galactic conformity,'' which was first noted by \citet{weinm06}. They used a large galaxy group catalog (with $10^{12}M_{\sun}<\rm{M}_{\rm{halo}}<10^{15}M_{\sun}$) constructed from the Sloan Digital Sky Survey (SDSS) and found that, for similar halo masses, early-type central galaxies have comparatively higher fractions of early-type quenched satellite galaxies around them, while late-type centrals tended to be surrounded by late-type satellites.
Several studies on conformity \citep[e.g.,][among others]{treyer18,berti17,kawin16,hartley15} have followed, but they have mostly focused on lower halo mass clusters and groups.

~~~Notably, it has been suggested by \citet{knob15}, in their study of conformity in the SDSS sample, that the environmentally driven quenching process in larger groups and clusters could essentially be  the same process as that which drives the mass-quenching. 
We may consider the effect of conformity as a "boosting of the quenching
effect of the environment  if the central is quenched," namely, an additional (mass-)quenching term that acts on the more massive SF satellites approaching the inner regions of clusters with passive central galaxies. One candidate for producing this boosting of the quenching effect are active galactic nuclei (AGN).
The first observational evidence for a higher number of AGN in massive galaxies around passive (non-SF) centrals was found by \citet{kauf15} when studying the SDSS sample. More recently, \citet{poggianti17} observed a high incidence of AGN among stripped jellyfish galaxies. They interpreted this as possibly due to ram pressure causing gas to flow toward the center of stripped galaxies and triggering AGN activity.


~~~To further interpret the observations of \citet{poggianti17},  the connection between ram pressure and AGN activity was analyzed by \citet{ricarte20} in the state-of-the-art ROMULUS\,C cosmological simulation \citep{tremmel19} of a $10^{14} M_{\odot}$ cluster, one of the highest resolution clusters ever simulated. They found that
for more massive galaxies, accretion onto the black hole is enhanced during pericentric passage. In one of their case study, RPS and AGN appear to work in synergy to shut down star formation. While ram pressure compresses the gas and helps fuel the
AGN, the AGN feedback acts on the (cold) gas deepest in the galaxy's potential well, which is the most difficult gas for RPS to remove. AGN feedback can make this gas more susceptible to stripping by heating and driving winds that radially redistribute this gas to the outer parts of the galaxy.   This suggests that AGN feedback may play a role in rapidly quenching (massive) galaxies,
when these approach the cluster center.

~~~ \citet[][M19a in the following]{maier19a} used the location of galaxies in the projected velocity versus position phase-space (phase-space diagram) to separate their studied cluster sample into a region of objects accreted many Gyr ago
and a region of recently accreted and infalling galaxies.
They found that SF cluster and field galaxies show similar
specific SFRs in a given stellar mass bin. 
On the other hand, in studying the mass-metallicity relation (MZR), M19a found that the gas phase metallicities of galaxies in the stellar mass complete sample with $\rm{log(M_{*}/M_{\odot})} \ge 10$ are displaced to higher values inside the virial radius ($R<R_{200}$) when compared with  galaxies with similar masses at larger clustercentric radii and field galaxies at similar redshifts.
The comparison with bathtub metallicity-SFR-mass model predictions with inflowing gas \citep{lilly13} indicates a slow-quenching scenario in which strangulation is initiated when galaxies pass $R \sim R_{200}$ by  stopping the  inflow of pristine gas which would otherwise dilute the interstellar medium (ISM). Their  SFRs are hardly affected for a period of time 
because these galaxies consume the available  gas in the disk \citep[see also][]{maier16,ciocan20}.

~~~Thus, in the slow-then-rapid quenching scenario advocated by M19a, strangulation (slow quenching) starts when galaxies enter the $R_{200}$ of massive clusters because the gas inflow has been stopped, but the galaxies remain SF, consuming the available cold gas. After the slowly quenching galaxies travel for about $1-2$\,Gyr toward the inner regions of the cluster (entering $R_{500}$), the RPS becomes larger and can possibly trigger AGN, which will completely (rapidly) quench galaxies (together with the RPS), as we want to explore in the present study.
The ``200'' and ``500'' subscripts refer to the radius where the average density is 200 and 500 times the critical density of the Universe at the cluster redshift, and $R_{200}$ roughly corresponds to the virial radius of the cluster. 
Interestingly, using the Yonsei Zoom-in Cluster Simulation (YZiCS),  \citet{rhee20} recently found that quenching remains gentle for roughly 2\,Gyr for satellite galaxies because gas loss happens primarily on hot and neutral gases, and then quenching becomes more dramatic,
because RPS is strongest near the cluster center. This is in agreement with the findings of M19a.

~~~Star formation in brightest cluster galaxies (BCGs) has been identified in several LoCuSS \citep[e.g.,][]{smith10}
clusters by using different star-formation tracers. The most plausible candidate for the gas fueling star-formation in these BCGs is radiatively cooled intracluster medium (ICM) plasma, with the cooling time for the ICM gas in the very core of the cluster ($R \lessapprox 30$\,kpc) being considerably shorter than the Hubble time at the redshift of the cluster (cool-core cluster).
We want to extend the studies of M19a by exploring galactic conformity manifested in  differences of the fractions of SF galaxies and quenched galaxies around active (with star-formation activity) and passive (without star-formation activity) BCGs.
Throughout the paper, we denote satellite galaxies in clusters with active BCGs as ABCG galaxies, and satellite galaxies in clusters with passive BCGs as PBCG galaxies.


~~~  The paper is structured as follows. In Sect.\,2, we
present the selection of the sample of 18 clusters for this study with suitable Hectospec spectroscopy.
We briefly recapitulate the selection of type-2 AGN, SF, and non-SF galaxies for the few clusters studied in M19a, which is now done for the entire sample of 18 clusters.
In Sect.\,3, we present our observational results.  The MZR of ABCG and PBCG cluster galaxies inside $R_{200}$ and $R_{500}$ is compared to the MZR of field galaxies at similar redshifts. 
We explore the phase-space diagram  for cluster objects classified as type-2 AGN, SF, and non-SF galaxies, as well as recently quenched galaxies (RQGs), and we compute the fractions of AGN, RQGs, and SF galaxies at $R<R_{500}$ around active and passive BCGs. The study of metallicities and of the fraction of AGN and RQGs reveals  that strangulation and rapid quenching act on PBCGs in the inner regions of clusters.
In Sect.\,4, we discuss the comparison of the observational results with theoretical simulations.
Finally, in Sect.\,\ref{sec:summary}, we summarize and discuss our conclusions.
As in M19a, a concordance cosmology with $\rm{H}_{0}=70$ \kmsmpc,
$\Omega_{0}=0.25$, $\Omega_{\Lambda}=0.75$ is used throughout this
paper.  
We assume a Salpeter \citep{salp55} initial mass function for all derived stellar masses.


\section{Data and measurements}

\subsection{Cluster galaxies from LoCuSS and sample selection}
~~~ Local Cluster Substructure Survey (LoCuSS)
is a multiwavelength survey of X-ray luminous galaxy clusters at $0.15<z<0.3$ \citep{smith10} drawn from the ROSAT All Sky Survey cluster catalog. These clusters benefit from a rich data set, including Subaru optical imaging, Spitzer 24\,$\mu$m maps, Herschel 100-500\,$\mu$m maps, Chandra and XMM X-ray data, GALEX UV (ultraviolet) data, and near-infrared imaging. %
 Arizona Cluster Redshift Survey (ACReS)
is a large spectroscopic survey using Hectospec/MMT following up 30 massive clusters of LoCuSS. The main strengths of ACReS are its wide field of view ($\sim 1$ degree diameter), which reaches well into the infall regions (up to about three virial radii) of the clusters, and the careful target selection based on J and K near-infrared imaging, which provides an unbiased, mass-complete sample of cluster galaxies, down to  $\rm{log(M/M_{\odot})} \sim 10$ (see more details in M19a).


~~~A spectral coverage including the H$\alpha$ and [NII]$\lambda6584$ emission lines (ELs) to identify AGN is required for our studies. Therefore, we excluded six of the 30 LoCuSS clusters with $z>0.26$, where the [NII] EL  was redshifted  out of the observed ACReS spectrum range ($4000-8300$\AA).
Four additional clusters that do not have the photometry  needed to determine stellar masses, and  two merging clusters (A1914 and A115, each with two BCGs) are also excluded. This results in a sample of 18 clusters, as shown in Table\,\ref{tab:20LoCuSS}: nine with an active BCG and nine with a passive BCG, making up a suitable sample for our galactic conformity studies. We note that six out of these 18 clusters were already studied by M19a, as indicated in Table\,\ref{tab:20LoCuSS}.

~~~The nine active BCGs have spectra with H$\alpha$ in emission  (due to star formation and AGN activity), and they are also listed as ``H$\alpha$-detected BCGs'' in Table\,2 of \citet{rawle12}.
 These authors showed that the majority of the BCGs with H$\alpha$ in emission in our sample are dominated by the stellar component, showing a  decreasing flux through the Spitzer-IRAC bands, as expected for a stellar-dominated spectral energy distribution \citep[e.g.,][]{egami06}. Only one of our studied nine BCGs with H$\alpha$ in emission, in cluster Z2089, shows a higher contribution from an AGN, but its SFR derived from comparing different infrared bands is still quite high \citep[see][]{rawle12}.
Passive BCGs have absorption line spectra (belonging to the nonSFgals sample, see below). The core entropies $K_{0}$ of the clusters are listed in  Table 5 of \citet{cavagnolo09}. The nine LoCuSS clusters with active  BCGs have $K_{0}<25keVcm^{2}$, classifying them as cool-core clusters, while all nine LoCuSS clusters with passive BCGs have $K_{0}>55keVcm^{2}$, classifying them as non cool-core clusters.

\begin{table*}
\caption{Eighteen clusters of this study:
  mean redshift of observed cluster members, coordinates, radii $R_{200}$ and cluster masses $M_{200}$ \citep[as reported in][]{haines13}. The last three columns indicate the number of observed cluster members, SF objects and galaxies with O/H measured in each cluster, for galaxies with $10\leq \rm{log(M/M_{\odot})}\leq 11$.}
\label{tab:20LoCuSS}
\begin{tabular}{ccccccccc}
\hline      
Cluster & z &   RA    &    DEC & $R_{200}$ & $M_{200}$             &  Cluster                     &   SF cluster   &  Members  \\
name    &   &         &        & (Mpc)    &  ($10^{14}M_{\odot}$) & members   &   members     &  with O/H  \\ 
\hline\hline
&&&PBCGs&&&&&\\
\hline
A586  & 0.1707 & 07:32:20.42  & +31:37:58.8   & 1.63  & 4.98 &  177 &   43   &  17  \\
A665  & 0.1827 & 08:30:57.36  & +65:51:14.4   & 2.19  & 12.04&  232 &   41   &  21  \\
A1689$^{a}$ & 0.1851 & 13:11:29.45  & $-$01:20:28.3 & 2.13 & 11.01 &  292 &   46  &  32  \\ 
A963$^{a}$  & 0.2043 & 10:17:01.20  & +39:01:44.4   & 1.81 & 6.75  &  333 &   75  &  45  \\ 
A2219 & 0.2257 & 16:40:22.56  & +46:42:21.6   & 2.77  & 24.30&  254 &   25   &  21  \\
Z1693 & 0.2261 & 08:25:57.84  & +04:14:47.5   & 1.58  & 4.53 &  235 &   48   &  31  \\
A267  & 0.2275 & 01:52:48.72  & +01:01:08.4   & 2.84 & 26.16 &  62   &   15  &   10  \\ 
A1763$^{a}$ & 0.2323 & 13:35:16.32  & +40:59:45.6   & 1.85 & 7.18  &  233 &   63   &  43  \\
A68   & 0.2510 & 00:37:06.84  & +09:09:24.3   & 1.39  & 3.10 &  143 &   29   &  20  \\
\hline
&&Total & Number& & &1961& 385& 240\\
\hline
\hline
&&&ABCGs&&&&&\\
\hline      
RXJ1720$^{a}$& 0.1599& 17:20:10.14  & +26:37:30.9   & 2.02 & 9.40  &  219 &    22   &  15  \\ 
A383  & 0.1887 & 02:48:03.42  & $-$03:31:45.1 & 1.52  & 3.98 &  137 &   27   &  20  \\
Z1883 & 0.1931 & 08:42:56.06  & +29:27:25.7   & 2.34  &14.53 &  127 &   27   &  15  \\
A291  & 0.1955 & 02:01:43.11  & $-$02:11:48.1 & 1.32  & 2.60 &  84  &   17   &  8   \\
A2390$^{a}$ & 0.2291 & 21:53:36.72  & +17:41:31.2   & 2.21 & 12.36 &  292 &   70   &  48  \\ 
RXJ2129& 0.2337& 21:29:40.02  & +00:05:20.9   & 1.73  & 5.88 &  195 &   38   &  19  \\
Z2089 & 0.2344 & 09:00:36.86  & +20:53:40.0   & 1.52  & 4.01 &  105 &   21   &  16  \\ 
A1835$^{a}$ & 0.2520 & 14:01:02.40  & +02:52:55.2   & 2.27 & 13.34 &  336 &   84  &  45  \\ 
Z348  & 0.2526 & 01:06:49.50  & +01:03:22.1   & 1.15  & 1.75 &  127 &   35   &  17  \\ 
\hline
&&Total & Number& & &1622& 341& 203\\
\hline
\end{tabular}
\\
\tablefoottext{a}{These six clusters were studied also by M19a.}\\
\end{table*}

\subsection{Star-forming galaxies, AGN, and recently quenched galaxies}
\label{sec:type2AGNs}


\begin{figure}[ht]
\includegraphics[width=7.5cm,angle=270,clip=true]{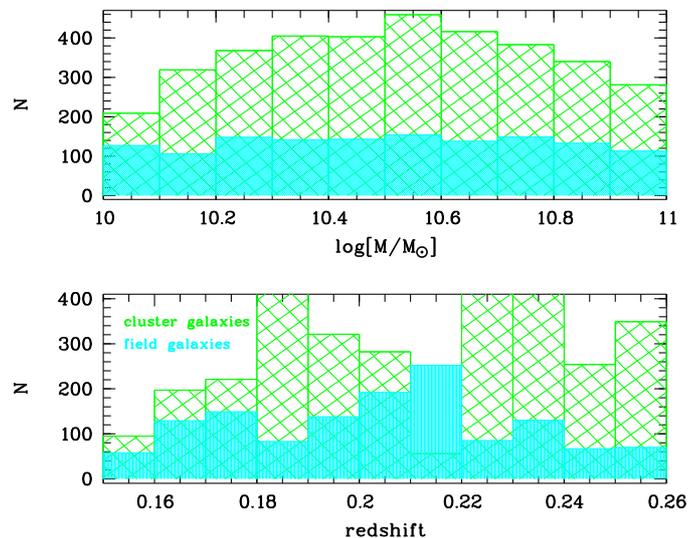}
\caption
{
\label{fig:zMass} 
\footnotesize 
Distribution of redshift and stellar mass of the mass-complete LoCuSS cluster sample (green)
and of the mass-complete field control sample (cyan).
}
\end{figure}

~~~The selection of cluster members and of a field control sample at $0.15<z<0.26$ from the diagram showing the spectroscopic redshift (derived from Hectospec) versus projected clustercentric radius was done using the isodensity contours (i.e., ``caustics'') as described in M19a; the field control sample has the same redshift and stellar mass range as the cluster galaxy sample (see Fig.\,\ref{fig:zMass}), and was observed at the same time with the same instrument Hectospec.
The $R_{200}$ and $R_{500}$ radii of each cluster were estimated by fitting deprojected gas and temperature profiles to the annular spectra of each cluster from the Chandra X-ray images, to determine the gravitational mass profile \citep{sanderson09}.
The ``200'' (``500'') subscript refers to the radius where the average density is 200 (500) times the critical density of the Universe at the cluster redshift, and $R_{200}$ roughly corresponds to the virial radius of the cluster. 
From the sample of about 3600 cluster galaxies in 18 clusters in the mass-complete sample  (see Table\,\ref{tab:20LoCuSS}), about 1800 objects lie at $R<R_{200}$, while the mass-complete field control sample contains about 1300 galaxies.

~~~For the determination of stellar masses of cluster galaxies from the available photometry, we refer to M19a. Gas metallicities (oxygen abundances) were derived from the fluxes of four ELs (of galaxies from the OHgals sample) using the following equation: $12+\rm{log}(O/H)=8.97-0.32 \times \rm{log}(([OIII]/\Hb)/([NII]/\Ha))$.
OHgals sample galaxies were selected as described in M19a based on signal-to-noise ratio (S/N) criteria for the four ELs and using the BPT \citep{bald81} diagram to exclude type-2 AGN.

~~~As in M19a, the \emph{SFgals} sample contains galaxies with a rest-frame $EW(H\alpha) >3$ \AA, a $S/N>5$ in  H$\alpha$ EL flux and log([NII]/H$\alpha)\leq -0.3$. The  \emph{AGN} sample of type-2 AGN contains objects with a rest-frame $EW(H\alpha) >6$ \AA,  a $S/N>5$ in  H$\alpha$ EL flux and log([NII]/H$\alpha)>-0.3$, following  Fig.\,6 of \citet{cidfern11}. Galaxies not belonging to the \emph{SFgals} or \emph{AGN} sample are classified as non-SF and belong to the \emph{nonSFgals} sample (see Table\,\ref{tab:subsamples}).

~~~Supplementary to the classification of M19a, we identified galaxies that have recently
undergone a cessation of their star formation, known as recently quenched galaxies (RQGs).
For their detection, we used two star-formation tracers: i) UV,  with blue $NUV-R \leq 4.5$ colors indicating star-formation, as shown by \citet{haines15} and ii) H$\alpha$, with an equivalent width of H$\alpha$ less than 3\AA\, indicating no recent star-formation in the last 10\,Myr (belonging to the \emph{nonSFgals} sample). 
If the star formation in a galaxy is shut down quickly, the O stars producing H$\alpha$ will die within 10\,Myr and the galaxy will not appear as SF in the  H$\alpha$ tracer anymore. On the other hand, the UV emission traces O- and B-stars, where the latter live for typically $\sim 0.1$Gyr. This implies that galaxies with UV emission but no H$\alpha$ stopped forming stars recently, namely, less than  $\sim 0.1$Gyr ago.

\begin{table*}
  \caption{Number of Mhigh objects in different subsamples of cluster galaxies with $R<R_{500}$ in clusters with passive BCGs (PBCG galaxies) and active BCGs (ABCG galaxies). See Sect.\,\ref{sec:type2AGNs} for details of the selection of the subsamples.}
\label{tab:subsamples}
\begin{tabular}{ccccccc}
\hline\hline      
Cluster sample& PBCG galaxies&&&&&\\
&Parent sample &   $SFgals$   &   $nonSFgals$ & $AGN$  & $OHgals$ & RQGs\\
\hline      
$\rm{10.7\leq log(M/M_{\odot})}\leq 11$ & 205 & 6 & 183 & 16 & 2 & 19 \\
\hline\hline      
Cluster sample& ABCG galaxies&&&&&\\
&Parent sample &   $SFgals$   &   $nonSFgals$ & $AGN$  & $OHgals$ & RQGs \\
\hline      
$\rm{10.7\leq log(M/M_{\odot})}\leq 11$  & 118 & 12 & 101 & 5 & 8 & 2 \\
\hline      
\hline
\end{tabular}
\\
\end{table*}


\section{Results: Strangulation and rapid quenching in clusters with passive BCGs}
\label{sec:results}

~~~To study galactic conformity effects, we concentrate on the small fraction of galaxies which enter $R_{200}$ and are still forming stars; in particular, galaxies that show ELs provide an opportunity for  their gas metallicities to be measured.
While the contamination by interlopers increases with clustercentric radius to more than 50\% up to $R = 3R_{200}$ \citep[see Fig.\,6 in][]{rhee17},  we can reduce the fraction of interlopers in our cluster sample to about 15\% \citep[as shown by][]{haines15}  by concentrating on $R<R_{200}$. These interlopers are galaxies in the infall regions projected along the line-of-sight. They are gravitationally bound to the cluster, and will eventually pass within $R_{200}$, but had not yet done so at the time of observation. 

~~~No massive EL satellite galaxy with a stellar mass of $\rm{log(M/M}_{\odot})>11$  has survived and continued to form stars at $R<R_{200}$  in our sample of 18 clusters. Therefore, for a meaningful investigation, we concentrate our analysis on the $10 \le log(M/M_{\odot})\le 11$ sample, and divide it for our analysis into two mass bins: galaxies with Mlow ($10 \le log(M/M_{\odot})< 10.7$, namely: $1 \cdot 10^{10} \le M/M_{\odot} < 5 \cdot 10^{10}$); and galaxies with Mhigh ($10.7 \le log(M/M_{\odot}) \le 11$, namely: $5\cdot 10^{10} \le M/M_{\odot} \le 10\cdot 10^{10}$).
%

\subsection{Strangulation revealed by enhanced metallicities of PBCG cluster galaxies inside $R_{200}$}
\label{sect:MZR}
\begin{figure*}[ht]
\includegraphics[width=12cm,angle=270,clip=true]{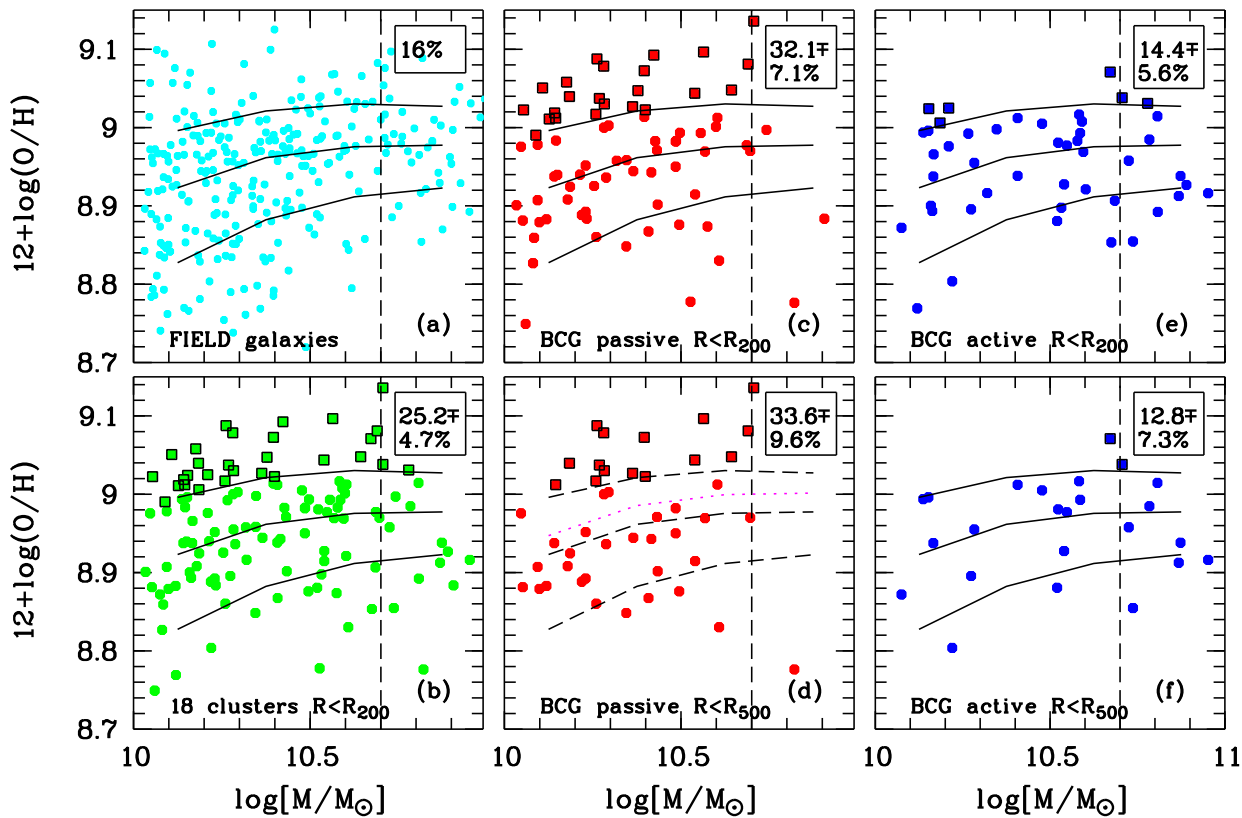} 
\caption
{
\label{fig:MassOH} 
\footnotesize 
MZR of galaxies with $0.15<z<0.26$: field galaxies (panel a),  cluster galaxies at $R<R_{200}$ (green symbols in panel b), PBCG cluster galaxies  (red symbols in panels c and d) and  ABCG cluster galaxies (blue symbols in panels e and f). 
For the coeval field galaxies in panel (a) individual measurements are depicted as cyan symbols, while we show 16th and 84th percentiles, and the medians (50th percentiles) of the distribution of O/H values as black lines which we repeat in every panel for comparison to the cluster galaxies.%
The percentage numbers in  each panel indicate the fraction of $10 \le log(M/M_{\odot})\le 11$ objects with high metallicities, above the 84th percentiles (upper black line) of the field galaxies; the symbols for these high O/H cluster galaxies are additionally framed by black squares. The vertical dashed line shows the division between the Mlow and Mhigh sample. The magenta dotted line in panel (d) indicates a gap in the PBCGs metallicity distribution between high metallicity and low metallicity objects.
}
\end{figure*}

\begin{figure*}[ht]  
\includegraphics[width=4.5cm,angle=270,clip=true]{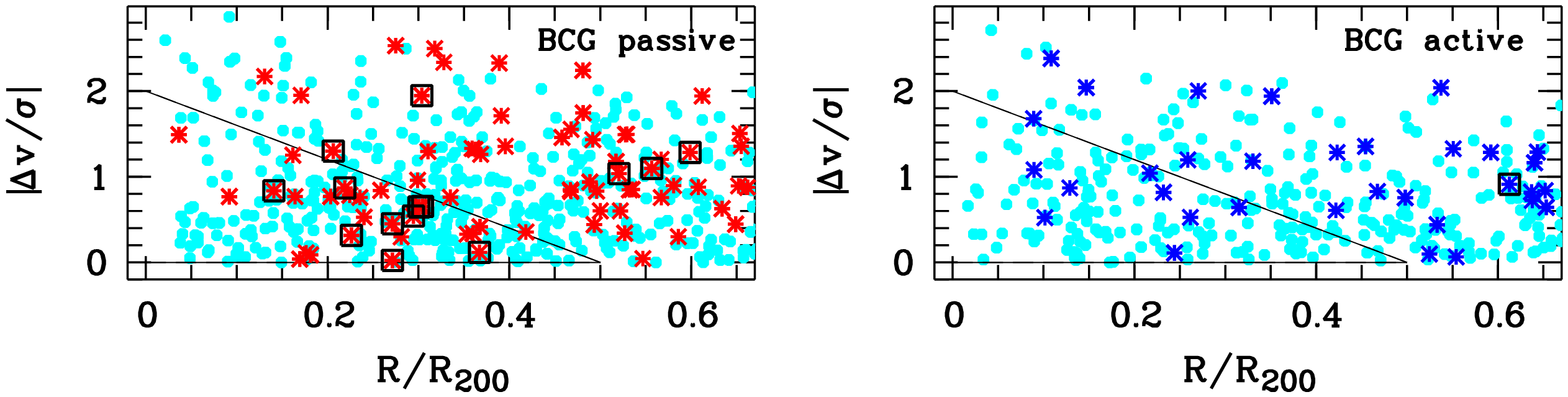}
\caption
{
\label{fig:PhaseSpLoCuSSMlow} 
\footnotesize 
Phase-space diagram for Mlow galaxies at $R<R_{500}$ for nine LoCuSS clusters with passive BCGs (left panels) and nine LoCuSS clusters with active BCGs (right panels). Non-SF galaxies (nonSFgals sample) are shown as cyan symbols, 
PBCG and ABCG SF galaxies are shown as red and blue asterisks, while the symbols for galaxies with higher O/H metallicities are additionally framed by black squares.
The triangle in the lower left-hand corners shows the innermost part of the virialized region containing less than 5\% interlopers (due to projection effects), as derived by \citet{rhee17} using cosmological hydrodynamic simulations of groups and clusters.
}
\end{figure*}


~~~ M19a found indications for strangulation based on enhanced gas phase metallicities of galaxies with $\rm{log(M_{*}/M_{\odot})} \ge 10$ observed inside $R_{200}$ of seven LoCuSS clusters. We now consider the larger sample of cluster galaxies (from 18 LoCuSS clusters) which entered $R_{200}$ and which survived the cluster environment and remained SF.
The MZR of the SF cluster galaxies with enough ELs observed to derive metallicities, the OHgals sample described in Sect.\,\ref{sec:type2AGNs}, and the MZR of the field control sample galaxies at similar redshifts are shown in Fig.\,\ref{fig:MassOH}.
With a higher number of LoCuSS clusters, we can now confirm the tentative finding of M19a of higher metallicities among SF galaxies in the virialized regions of the clusters ($R<R_{200}$, see panel b in Fig.\,\ref{fig:MassOH}), compared to our coeval field sample  (panel a in Fig.\,\ref{fig:MassOH}).

~~~We use the following to denote galaxies with high or enhanced metallicities as those objects above the 84th percentile of the distribution of field galaxies in the MZR diagram, shown as the upper solid black line in the panels in Fig.\,\ref{fig:MassOH}.
A gap in the distribution of metallicities of PBCG galaxies (panel d in Fig.\,\ref{fig:MassOH}) can be seen, indicated by a magenta dotted line. This is a reason why we do not plot median values of metallicities of cluster galaxies, but rather concentrate on the comparison of the fraction of galaxies with enhanced O/Hs, which is not affected by this gap.
By comparing the fraction of galaxies with high metallicities (symbols framed by black squares in Fig.\,\ref{fig:MassOH}) in clusters and coeval field regions, we see that the effect of enhanced O/Hs seems to be entirely due to the cluster galaxies around passive BCGs. On the one hand, 16\% of field galaxies lie above the upper solid black line by definition and a similar percentage of $14.4\pm5.6$\%  of ABCG cluster galaxies at $R<R_{200}$
lie in this region, as shown in panel (e). On the other hand, a much higher percentage of 32.1\% PBCG cluster galaxies  in the OHgals sample at $R<R_{200}$  have high metallicities (panel c). At $R<R_{500}$, the difference between the high metallicity fraction of PBCG galaxies (33.6\%, panel d) and  the high metallicity fraction of ABCG galaxies (12.8\%, panel f) becomes even larger, albeit with larger uncertainties. Thus, while the metallicities of ABCG SF galaxies  at $R<R_{200}$ follow a similar MZR distribution as the bulk of field galaxies, a higher fraction of PBCG SF galaxies at $R<R_{200}$ show high metallicities compared to the population of coeval SF field galaxies.
This is in agreement with the M19a strangulation scenario for galaxies entering $R_{200}$, stopping the pristine gas inflow which would otherwise dilute their interstellar medium and would maintain their metallicities at values similar to those of field galaxies. For a more detailed explanation, we refer to Sect.\,4 of M19a and the equilibrium metallicity equation described there, which depicts how stopping the inflow of gas can increase gas metallicities.
The new result here is that the strangulation is revealed by the higher metallicities of SF cluster galaxies entering $R_{200}$ of clusters with passive BCGs, but not by ABCG galaxies.

\begin{figure*}[ht]  
\includegraphics[width=12cm,angle=270,clip=true]{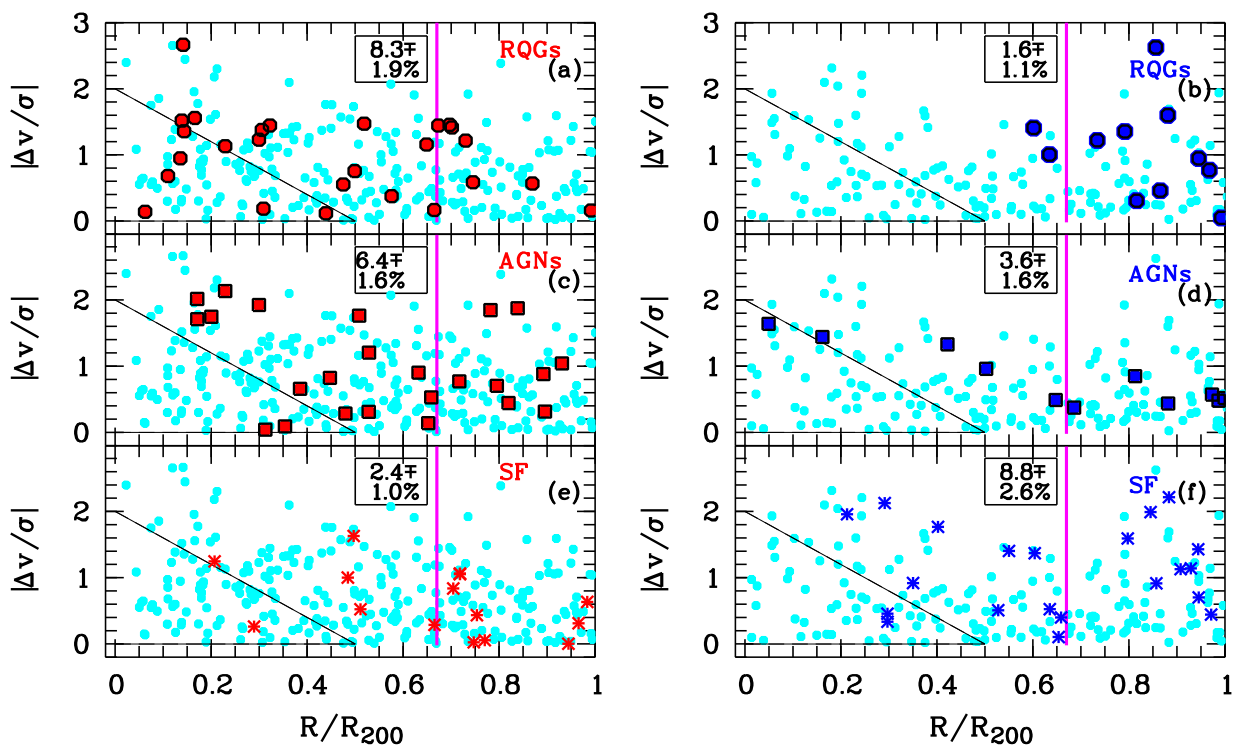}
\caption
{
\label{fig:PhaseSpLoCuSS} 
\footnotesize 
Phase-space diagram at $R<R_{200}$ for nine LoCuSS clusters with passive BCGs (left panels) and nine LoCuSS clusters with active BCGs (right panels). Objects in the Mhigh mass bin are shown, with non-SF galaxies (nonSFgals sample) shown as cyan symbols in all panels, with a higher number of nonSFgals objects with spectroscopy in the nine clusters with passive BCGs (left panels). SF galaxies with star formation traced by their H$\alpha$ EL (SFgals sample)  are shown as asterisks in the lower panels (e) and (f), type-2 AGN are shown as filled squares in the panels (c) and (d),  and galaxies which have recently undergone the cessation of their star formation (RQGs), with a SF UV component, but a lack of H$\alpha$, are shown by filled circles in the upper panels (a) and (b). The triangle in the lower left-hand corners shows the innermost part of the virialized region as described in the caption of Fig.\,\ref{fig:PhaseSpLoCuSSMlow}. The magenta vertical line indicates $R_{500}$. The percentage numbers in each panel are the fractions, within $R_{500}$, of either SF galaxies, AGN, or RQGs; as indicated in the respective panel, relative to the total number of Mhigh objects inside $R_{500}$ in the entire spectroscopic sample of nine clusters with either active or passive BCGs, respectively.
}
\end{figure*}

~~~Another interesting new result is that ABCG ELGs at $R<R_{500}$, with metallicity values similar to the bulk of $0.15 < z <0.26$ field galaxies, are found up to  $\rm{log(M/M_{\odot})} \sim 11$ (panel f), while PBCG SF ELGs galaxies with $\rm{log(M/M_{\odot})} \ge 10.7$ are almost absent at $R<R_{500}$  (panel e). Thus, there exists a threshold at $\rm{log(M/M_{\odot})} \sim 10.7$ above which ABCG ELGs at $R<R_{500}$ are found, but PBCG ELGs virtually disappear. The ELs and normal gas metallicities for Mhigh ABCG galaxies at $R<R_{500}$ with masses above the threshold indicate that these galaxies are still forming stars and that they evolve more slowly toward lower specific SFRs, that is to say,  their quenching is slower. On the other hand, Mhigh PBCG galaxies are quenched after entering $R<R_{500}$ and disappear (with one exception) from the SF population of ELGs, as shown in panel (d) of Fig.\,\ref{fig:MassOH}. Four additional SF Mhigh PBCG galaxies at $R<R_{500}$ are not part of the OHgals sample (see Table\,\ref{tab:subsamples}) because their EL are not sufficiently strong to measure their metallicity. They are still actively forming stars, but at reduced levels. The near absence of Mhigh PBCG ELGs galaxies at $R<R_{500}$ above the $\rm{log(M/M_{\odot})} \sim 10.7$  mass threshold indicates that these PBCG galaxies evolve faster toward lower specific SFRs, that is, they quench faster.

~~~The phase-space diagram (projected clustercentric
radius R versus line-of-sight velocity relative to the cluster redshift) provides valuable information on the accretion history of cluster galaxies \citep[e.g.,][]{haines15}. Figure\,\ref{fig:PhaseSpLoCuSSMlow} shows the stacked phase-space diagram for Mlow ABCG galaxies (right panel) and Mlow PBCG galaxies (left panel) at $R<R_{500}$, with the innermost part of the virialized region (triangle in the left hand corner) containing less than 5\% interlopers, as defined by \citet{rhee17}. Among Mlow SF galaxies entering this innermost region (triangle) of clusters with passive BCGs, more than one-third show high metallicities (red symbols framed by large black squares).
On the other hand, SF galaxies entering this innermost region in clusters with active BCGs exhibit lower metallicities, typical of the bulk of field galaxies (only one symbol framed by a black square in the right panel of Fig.\,\ref{fig:PhaseSpLoCuSSMlow}). Thus, the effect of high metallicities for PBCGs compared to ABCGs appears when galaxies pass $R<R_{200}$, as shown in Fig.\,\ref{fig:MassOH}; and it still exists for surviving SF galaxies entering the region at smaller clustercentric radii (triangle in Fig.\,\ref{fig:PhaseSpLoCuSSMlow}).

~~~Consequently, we find an indication for a strangulation scenario for $\rm{log(M/M_{\odot})} \ge 10$ PBCG galaxies entering $R<R_{200}$ and more rapid quenching of $\rm{log(M/M_{\odot})} \ge 10.7$ PBCG galaxies at $R<R_{500}$. This indicates a faster evolution toward low specific SFRs (faster quenching) for PBCG galaxies compared to ABCG galaxies. On the other hand, the metallicities and ELs of $10 \le \rm{log}(\rm{M}/\rm{M}_{\sun}) \le 11$  ABCG SF galaxies  are  less affected by environment on their way to the center of the cluster, and several Mhigh ABCG ELGs with metallicities typical of the coeval field population are still found at $R<R_{500}$. The existence of a threshold of $\rm{log(M/M_{\odot})} \sim 10.7$ above which ABCG ELGs at $R<R_{500}$ are found, but PBCG ELGs virtually disappear, points toward a faster evolution towards quenching for Mhigh PBCG galaxies at $R<R_{500}$. We discuss the stronger quenching of Mhigh PBCG galaxies at $R<R_{500}$ in the next section.

\subsection{Rapid quenching revealed by the increased fraction of AGN and RQGs inside $R_{500}$ around passive BCGs}
\label{sect:AGNsRapQ}

~~~To further explore the reason for the differences in properties of PBCGs and ABCGs seen in Fig.\,\ref{fig:MassOH}, we now  concentrate on Mhigh PBCGs and ABCGs at $R<R_{500}$, with stellar masses above the mass threshold found in the previous section. We want to explore whether a synergy between two effects (AGN and RPS) is responsible for quenching satellite galaxies approaching the inner regions of clusters. Based on their study of conformity in the SDSS sample, \citet{knob15} supposed that the environmentally driven quenching process in clusters and the mass-quenching are related. Figure 13 in \citet{peng10} depicts the mass functions in high density regions for mass-quenched satellites (likely due to AGN) and environment-quenched satellites (likely due to RPS), with environment-quenching dominating at lower stellar masses. With increasing stellar mass, the mass-quenching becomes stronger reaching similar strengths as the environment-quenching at high stellar masses in high density regions. We focus therefore on the Mhigh PBCGs and ABCGs in the following in order to study whether the combination of the two quenching effects of comparable strength at these higher stellar masses produces a conformity signal.

~~~From the phase-space diagram analysis of seven LoCuSS clusters, M19a showed how the observed fraction of SF galaxies decreases inside the virial radius ($R<R_{200}$) compared to field galaxies and  outer regions of clusters  up to $3R_{200}$ (see Fig.\,7 in M19a). Here, we want to explore the population of surviving SF cluster galaxies in the inner regions of the clusters at $R<R_{500}$, by comparing Mhigh ABCG, and PBCG galaxies. In particular, we want to compare, respectively, the fractions of SF galaxies, RQGs and AGN (relative to the total number of spectroscopic confirmed members with Mhigh) in clusters with active BCGs and passive BCGs.

~~~Figure\,\ref{fig:PhaseSpLoCuSS} shows the stacked phase-space diagram for the $R<R_{200}$ massive (Mhigh) galaxies for the nine clusters with passive BCGs (left panels) and nine clusters with active BCGs (right panels). The innermost part of the virialized region that contains a percentage of interlopers that is less than 5\%  , as defined by \citet{rhee17}, is depicted as a triangle. Because the number of Mhigh SF galaxies in this innermost region (triangle) is very low, we focus our analysis of the fraction of SF galaxies, AGN and RQGs on objects within $R_{500}$ to get higher number statistics. By focusing on  $R<R_{500}$ instead of $R<R_{200}$ we get a reduction of the contamination of our cluster sample with interlopers to less than about 10\% \citep[following Fig.\,6 in][]{rhee17}, a contamination that increases with clustercentric radius: about 15\% at $R_{500}<R<R_{200}$, and larger than 15\% at $R>R_{200}$.  Additionally, to study galactic conformity effects, we must compare the fraction of satellite galaxies with masses similar or close to the masses of the BCGs ($\rm{log(M/M_{\odot})} \sim 11$ and higher); therefore, we focused the analysis of the fractions on the Mhigh LoCuSS sample.

~~~A feature of galactic conformity is revealed in the panels (e) and (f) of Fig.\,\ref{fig:PhaseSpLoCuSS}: the fraction $f_{SF}$ of Mhigh ABCG SFgals at $R<R_{500}$  is  higher ($8.8 \pm 2.6$\%) compared with $f_{SF}=2.4 \pm 1.0$\% of PBCG galaxies (see also Table\,\ref{tab:subsamples}). On the other hand,  a higher fraction of type-2 AGN in clusters with a passive BCG  ($f_{AGN}=6.4\pm1.6$\% at $R<R_{500}$) compared to clusters with an active BCG ($f_{AGN}=3.6\pm1.6$\%) is seen in panels (c) and (d) of Fig.\,\ref{fig:PhaseSpLoCuSS}. The error bars of the fractions given in Fig.\,\ref{fig:PhaseSpLoCuSS} are computed as described in detail in M19a, by weighting each galaxy by the inverse probability of the galaxy to have been observed spectroscopically, and then by using Poisson statistics with the formula given in M19a. We checked that excluding the two most massive clusters with passive BCGs (see Table\,\ref{tab:20LoCuSS}) virtually does not change the fraction of RQGs and AGN at $R<R_{500}$ for PBCGs, but it lowers the fraction of SF PBCGs at $R<R_{500}$, making the conformity effect even more pronounced.

~~~A striking effect is seen by comparing panels (a) and (b) of Fig.\,\ref{fig:PhaseSpLoCuSS}, which show the existence of numerous Mhigh  RQGs at all radii within PBCG clusters (including the core regions), but the complete absence of  Mhigh ABCG RQGs  at $R<0.5R_{200}$. At $R<R_{500}$, a much higher fraction of PBCG RQGs of $f_{RQG}=8.3\pm1.9$\%  is found, as compared to the very low fraction of $f_{RQG}=1.6\pm1.1$\%  for ABCG galaxies. A plausible explanation is that the recent quenching of these galaxies in clusters with passive BCGs is connected to the higher fraction of AGN at $R<R_{500}$ in clusters with passive BCGs (panel c of Fig.\,\ref{fig:PhaseSpLoCuSS}),  which possibly (and recently) quickly quenched some galaxies around passive BCGs.

~~~An increasing number of AGN toward the center of clusters with passive BCGs as found for LoCuSS galaxies can be interpreted, as shown by \citet{ricarte20} in their analysis of the ROMULUS\,C cosmological simulation, as being due to the increasing  RPS toward the central regions of clusters that can
compress the gas and help fuel an AGN.
By heating and driving winds that radially redistribute the gas located deepest in the galaxy's potential well to the outer parts of the galaxy, AGN feedback can make this gas more susceptible to stripping. In this case, RPS and AGN work in synergy to shut down star formation. This suggests that AGN feedback may play a role in rapidly quenching massive galaxies, when these approach the centers of clusters with passive BCGs.


\section{Comparison with theoretical simulations}
\label{delayrapid}


~~~ We searched the literature for numerical simulations of massive clusters to compare the observed and predicted positions of SF galaxies and passive galaxies in phase-space diagrams from recent cosmological hydrodynamic simulations of clusters that include halos with $M_{vir}>10^{14.5}M_{\odot}$, such as our studied 18 LoCuSS clusters (see Table\,\ref{tab:20LoCuSS}). Apart from the ROMULUS\,C simulation discussed above,  we looked at the phase-space distributions of SF and quenched galaxies in simulated clusters from Magneticum Pathfinder \citep[see, e.g.,][]{lotz19} and from the IllustrisTNG simulations \citep[see, e.g.,][]{joshi20}.

~~~Magneticum Pathfinder is a set of large-scale smoothed-particle hydrodynamic simulations that employ a mesh-free Lagrangian method aimed at following structure formation on cosmological scales, with open access to many features \citep{ragagn17}. Two boxes are available through open access, Box2/hr ($352\,(\rm{Mpc\,h}^{-1})^{3}$) and Box2b/hr ($640\,(\rm{Mpc\, h}^{-1})^{3}$). We concentrate on the larger Box2b/hr, which allows for a more extended
statistical sample, especially regarding high-mass clusters. For more details on the simulations see, for instance, \citet{lotz19}.
We found 69 clusters with similar masses as our observed LoCuSS clusters, in the Box2b/hr snap\_31 ($z=0.25$) snapshot. However, for all these clusters, the BCGs are forming stars. Additionally, only $1-3$ galaxies inside $R_{500}$ are forming stars, that is, $\sim 0.5-1.5$\%, which is a much smaller percentage as we see in our observed LoCuSS clusters. Due to these features of the Magneticum Pathfinder simulations, (i) only SF BCGs and (ii) a too low percentage of SF galaxies at $R<R_{500}$, given  the overly strong quenching mechanism, these simulations are not well suited to be compared to our observed clusters.


\begin{table}
  \caption{Eighteen $z=0.2$ massive clusters from the IllustrisTNG-300-1 simulations: Id of IllustrisTNG-300-1 cluster, with radii, $R_{200}$, cluster masses, $\rm{log}(M_{halo}/M_{\odot})$, number of galaxies, $N_{gal}$, and number of SF galaxies, $N_{SFgals}$, in the sample of galaxies with $10 \le \rm{log(M/M_{\odot})} \le 11$ at $R<R_{200}$.
}
\label{tab:18TNG}
\begin{tabular}{ccccc}
\hline\hline      
TNG300  &$R_{200}$ & $\rm{log}(M_{halo}/M_{\odot})$ &$N_{gal}$ & $N_{SFgals}$ \\ 
cluster &(Mpc) &   & & \\ 
\hline
&BCG &passive& \\
\hline
CL2  &  2.15 & 14.95 & 55 & 1  \\ 
CL4  &  2.00 & 14.85 & 61 & 0  \\ 
CL7  &  1.90 & 14.74 & 59 & 0  \\ 
CL11 &  1.81 & 14.69 & 44 & 1  \\ 
CL19 &  1.70 & 14.56 & 33 & 3  \\ 
CL22 &  1.69 & 14.60 & 30 & 0  \\ 
CL28 &  1.66 & 14.55 & 39 & 0  \\ 
CL30 &  1.53 & 14.52 & 18 & 0  \\ 
CL31 &  1.55 & 14.52 & 31 & 1  \\ 
\hline
&Total &number& 370&6\\
\hline
&BCG &active&& \\
\hline
CL13&  1.81 & 14.60  & 49 & 2  \\ 
CL14&  1.74 & 14.64  & 35 & 2  \\ 
CL16&  1.66 & 14.60  & 31 & 1  \\ 
CL17&  1.74 & 14.60  & 33 & 0  \\ 
CL18&  1.75 & 14.59  & 35 & 3  \\ 
CL20&  1.73 & 14.58  & 39 & 1  \\ 
CL23&  1.74 & 14.58  & 30 & 3  \\ 
CL24&  1.56 & 14.51  & 44 & 0  \\ 
CL26&  1.67 & 14.52  & 26 & 0  \\ 
\hline
&Total &number& 322&12\\
\hline
\hline
\end{tabular}
\end{table}


~~~IllustrisTNG is a suite of cosmological simulations covering three volumes of $35\,(\rm{Mpc\,h}^{-1})^{3}$, $75\,(\rm{Mpc\,h}^{-1})^{3}$ and $205\,(\rm{Mpc\,h}^{-1})^{3}$ (TNG50, TNG100, and TNG300, respectively). The simulations are run using the moving-mesh code AREPO \citep{springel10}. The IllustrisTNG models feedback from supernovae in the form of galactic winds, and from AGN in the form of a thermal energy injection during the high-accretion
mode and kinetic energy during the low-accretion mode. For more details on the simulations see, for instance, \citet{joshi20}.

From the public data access webpage\footnote{https://www.tng-project.org/data/} we selected the 18 most massive clusters in IllustrisTNG300-1 that do not show signs of merging at $z=0.2$, with masses of $\rm{M_{halo}} \sim 10^{14.5-15}M_{\odot}$,  as indicated in Table\,\ref{tab:18TNG}, comparable to our massive LoCuSS clusters. We concentrated on TNG300 clusters, because TNG100 and TNG50 contain only lower-mass clusters  ($\rm{M_{halo}} < 10^{14.6}M_{\odot}$), as described by \citet{joshi20}. Galaxies are identified as any subhalo with a non-zero total stellar mass.

The SFR of a galaxy is measured within twice the stellar half-mass radius and is the sum of the instantaneous SFRs of all gravitationally bound gas cells contained within it. Stellar masses $M_{*}$ are also measured within twice the stellar half-mass radius. Galaxies with a $log(SFR/M_{*}/Gyr^{-1})<-3$ are considered to be passive because they are not forming stars at a cosmologically significant rate, namely: $sSFR\ll \tau_{H}^{-1}$ \citep[see, e.g., Fig.\,4 in][]{maier09}.

The number of SF galaxies  inside $R_{200}$ is quite small in the TNG300-1 simulations, as shown in Table\,\ref{tab:18TNG}. The quenching mechanism seem to be much stronger in the IllustrisTNG simulations compared to the observed LoCuSS clusters \citep[cf. the high quenched fraction found in massive clusters in IllustrisTNG by][]{donnari21}.

In summary, the simulations of clusters of galaxies show an overly strong quenching mechanism as compared to the observations. New theoretical simulations with more realistic quenching mechanisms in massive clusters are needed to explore why stronger quenching processes are observed in our explored LoCuSS clusters for PBCG galaxies as compared to ABCG galaxies.


\section{Summary and discussion}
\label{sec:summary}

~~~This study of galactic conformity and quenching in clusters is based on Hectospec spectroscopy of  galaxies in 18 LoCuSS clusters at $0.15<z<0.26$. The main results can be summarized as follows:

~~~1. The metallicities of log(M/M$_{\sun}) \geq 10$ cluster galaxies at $R<R_{200}$ are enhanced compared to field galaxies of similar masses at similar redshifts $0.15<z<0.26$, confirming the tentative result from M19a, based on the analysis of seven LoCuSS clusters. The enhancement in metallicities is entirely due to the  cluster galaxies which remain SF at $R<R_{200}$ in clusters with passive BCGs (Fig.\,\ref{fig:MassOH}). This indicates that strangulation is initiated when surviving SF satellite galaxies pass $R_{200}$, but only in clusters with passive BCGs. 

~~~2.  ABCG galaxies with metallicity values similar to the bulk of field galaxies at similar redshifts are found inside $R_{200}$ up to stellar masses of log(M/M$_{\sun}) \sim 11$, while  Mhigh (log(M/M$_{\sun})  \geq 10.7$) PBCG galaxies with ELs enabling gas metallicity measurements are nearly absent inside $R_{200}$ and especially inside $R_{500}$ (Fig.\,\ref{fig:MassOH}). This indicates that surviving log(M/M$_{\sun}) \geq 10$ ABCG SF galaxies entering $R_{200}$ are less affected by environmental quenching processes compared to PBCG ELGs. Quenching processes are stronger for PBCG ELGs, especially for Mhigh galaxies at $R<R_{500}$.

~~~3. In exploring the location of lower mass (Mlow, log(M/M$_{\sun}) < 10.7$) galaxies in the phase-space diagram, we find higher metallicities in more than one-third of the SF galaxies entering the innermost (virialized) region around passive BCGs, while the surviving Mlow SF galaxies in the innermost regions around active BCGs do not show higher metallicities compared to field galaxies (Fig.\,\ref{fig:PhaseSpLoCuSSMlow}). This indicates that  surviving SF Mlow ABCG  galaxies are hardly affected by environment on their way to the center of the cluster, while Mlow PBCG galaxies suffer strangulation (traced by enhanced metallicities) when traveling from $R_{200}$ to more inner regions of clusters.

~~~4.  To further explore the stronger quenching of Mhigh PBCG galaxies, we used the location of Mhigh SF galaxies, RQGs, and AGN in the phase-space diagram to compare their fraction at $R<R_{500}$ around active and passive BCGs. We find a higher fraction of Mhigh ABCG SF galaxies at $R<R_{500}$ ($8.8\pm2.6$\%), compared to SF PBCGs ($2.4\pm1.0$\%), which is as expected with regard to galactic conformity (Fig.\,\ref{fig:PhaseSpLoCuSS}, panels e and f).

~~~5.  On the other hand, we find a higher fraction of AGN at $R<R_{500}$ in clusters with passive BCGs  ($6.4$\%) compared to clusters with active BCGs ($3.6$\%). The fraction of Mhigh RQGs at $R<R_{500}$ in clusters with passive BCGs  ($8.3\pm1.9$\%) is much higher compared to the fraction of RQGs in the ABCGs population ($1.6\pm1.1$\%) and we do not find any Mhigh ABCG RQGs at $R<0.5R_{200}$. We assume that more rapid quenching of Mhigh PBCG galaxies entering $R_{500}$ produces more Mhigh RQGs at $R<R_{500}$ due to AGN that recently (and possibly quickly) quenched some PBCG Mhigh galaxies.

~~~We conclude that we see a slow shutting-down of star formation (strangulation) of $M < 5\cdot 10^{10}M_{\odot} $
surviving SF satellite galaxies entering $R_{200}$ of clusters with passive BCGs, which is not seen for Mlow surviving SF satellite galaxies entering $R_{200}$ of clusters with active BCGs. These PBCG SF galaxies continue to form stars consuming the available gas in the disk.  PBCG galaxies with $M > 5\cdot 10^{10}M_{\odot} $ surviving to be SF at $R<R_{500}$ suffer a rapid quenching of star formation likely due to AGN triggered by the increasing RPS toward the center of clusters.  These AGN can rapidly quench and maintain quenched satellite galaxies. However, more observations of individual cluster galaxies using integral field units are needed to explore whether the additional star formation quenching term that acts on the surviving PBCG SF satellites entering the inner regions of  clusters is indeed caused by AGN triggered by the increasing ram pressure in the inner denser regions of these clusters.

~~~On the other hand,  we found that  $M  \ge 10^{10}M_{\odot} $ surviving SF satellite galaxies entering $R_{200}$ of clusters with active BCGs continue forming stars when traveling to the inner regions of the cluster. Such SF galaxies with masses up to $M  \sim 10^{11}M_{\odot} $ and with  metallicities typical of field galaxies at similar redshifts are observed in clusters with active BCGs.

~~~The observed galactic conformity in the LoCuSS sample implies that the activity in the active BCGs must be maintained over relatively long timescales ($\sim 1$\,Gyr or more), since we would not observe conformity at $R<R_{200}$ (or $R<R_{500}$) if the activity in the central BCG would  rapidly be switching on and off.  This is due to the fact that the numbers of active SF galaxies in the rest of the cluster vary only significantly on Gyr timescales,  because it takes about one Gyr for these galaxies to fall into the cluster (see discussion on crossing timescales of LoCuSS clusters in M19a). Additionally, the enhanced metallicities of satellites around passive BCGs imply slow quenching (strangulation) over $1-2$\,Gyr, as discussed in M19a.

~~~Many previous studies on conformity \citep[e.g.,][]{treyer18,berti17,kawin16,hartley15} have studied lower mass halos (clusters and groups) with a low number
of satellite galaxies per group or cluster. The masses of those lower mass halos were estimated from the summed stellar mass content of the galaxies themselves, which can lead to significant biases. In our present work, we studied an X-ray selected sample of massive LoCuSS clusters, enabling more robust halo mass estimates, which is important especially when exploring conformity effects at a given halo mass.

~~~Using the parlance of \citet{knob15}, the existence of "conformity" in the LoCuSS sample tells us that there is still an additional unknown "hidden common variable" that is affecting the quenching of both satellites and centrals. The difficulty is that some unknown  "hidden common variables" are almost unobservable, for instance, the time that has elapsed since some particular event such as cluster formation. It is possible to obtain some information on  these unobservable variables by "observing" them in simulations. However, as discussed above, the quenching mechanisms are too strong in the simulations as compared to the LoCuSS observations. Therefore, new theoretical simulations with more realistic quenching mechanisms in clusters are needed in order to further explore the reasons why an additional quenching effect is seen in PBCG in comparison to ABCG galaxies.


\begin{acknowledgements}
We would like to thank the anonymous referee for providing constructive comments and help in improving the manuscript. CPH acknowledges support from ANID through Fondecyt Regular 2021 project no. 1211909.
\end{acknowledgements}



\end{document}